\documentclass{elsart}
\usepackage{natbib}
\usepackage[dvips]{epsfig}
\begin{document}
\runauthor{E.E. Boos}
\begin{frontmatter}
\title{Top Quarks at Photon Colliders}
\author[Moscow]{Edward Boos\thanksref{Someone}}

\address[Moscow]{Institute of Nuclear Physics, Moscow State University}
\thanks[Someone]{e-mail: boos@theory.npi.msu.su}
\begin{abstract}
A review of results on top quark physics expected at the Photon Linear
Colliders is presented. 
\end{abstract}
\end{frontmatter}

\section{Introduction}
The top quark, with the mass
slightly less than the mass of the gold nucleus, is the heaviest 
elementary particle found so far.
The RUN1 result
for the top mass measurement  
by the Fermilab CDF and D0 
collaborations is
$M_t = 174.3\pm 3.2 (stat) \pm 4.0 (syst)$ (see \cite{dudko}).

The top decays much faster than is typical for a formation
of the strong bound states. So, the top provides, in principle, a very clean
source for fundamental information. 
All top quark couplings to gauge bosons and other quarks are uniquely fixed
in the SM by the gauge principle, the structure of generations  and
a requirement of the lowest
dimension of the interaction Lagrangian which lead to
a renormalizability
and unitarity of the SM as a quantum field gauge theory.

The top is heavy and up to now point like at the same time.
The top Yukawa coupling $\lambda_{t} = 2^{3/4}G_F^{1/2}m_{t}$ is
numerically very close  to unity, and it is not clear whether or not this
takes place due to some deep physical reason.
Because of unusual top properties  
one might expect deviations from the SM predictions to be more
likely in the top sector \cite{peccei}. 
Studies of the top may shed light on the origin of
the mechanism of EW symmetry breaking.   
Top quark physics will be a very important part of research programs
for all future hadron and lepton colliders. The $\gamma\gamma$
collider is of special interest because of a very clean
electromagnetic production mechanism with high rate (see the review
\cite{hewett}).

\section{Top pair production in $\gamma \gamma$ collisions}
The leading order $\gamma\gamma \to t \bar{t}$ cross section
is well known and has a maximum a maximum of about 420 $(++)$, 450 (unpolarized) and
550 $(+-)$ GeV.
This cross section is larger than the corresponding $e^+e^-$
cross section. 

\begin{figure*}[h!b]
\begin{center}
\mbox{\epsfxsize=14cm\epsfysize=12cm\epsffile{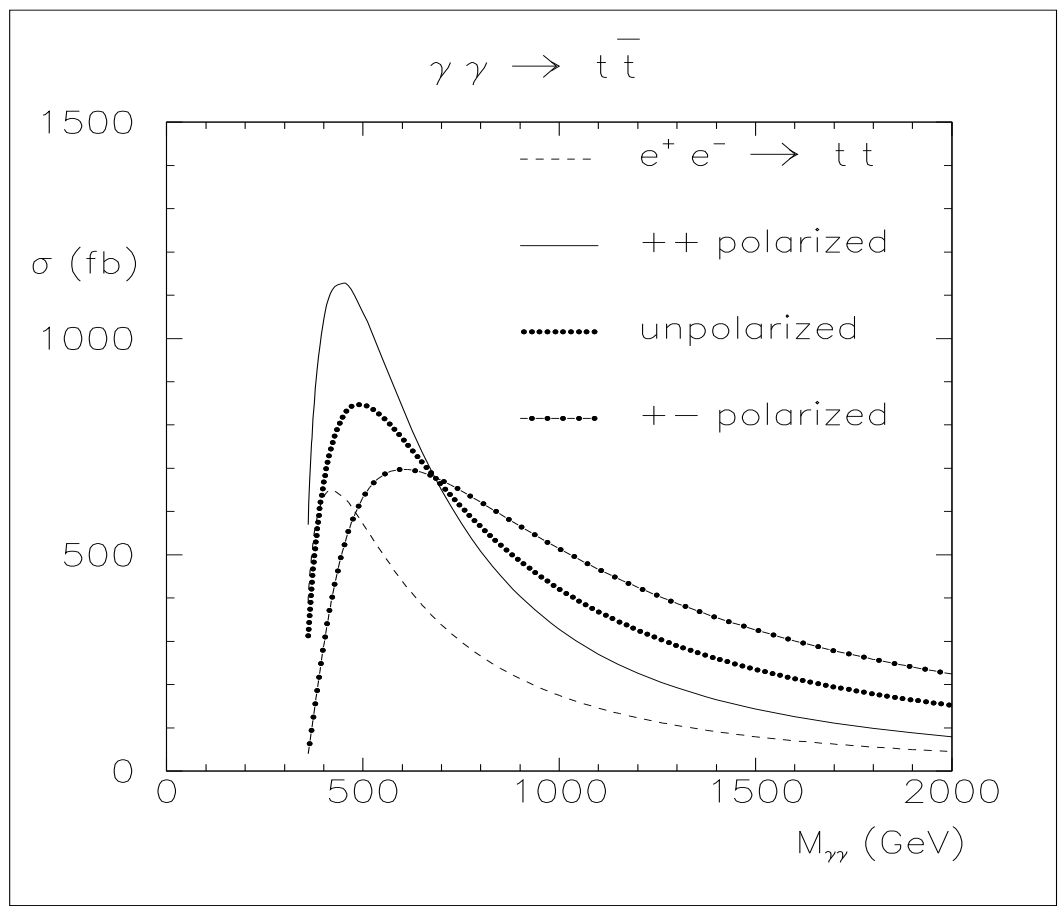}}
\end{center}
\caption{Top quark pair production to LO}
\label{fig:f1}
\end{figure*}

The $(++)=(--)$ helicity configuration ($J_{z}=0$) dominates at 
energies less than about 680 GeV, while the $(+-)=(-+)$ or $J_{z}=2$
configuration starts to dominate at higher energies.  

\begin{figure}[htbp]
\begin{center}
\centering\mbox{\epsfxsize=14cm\epsfysize=12cm\epsffile{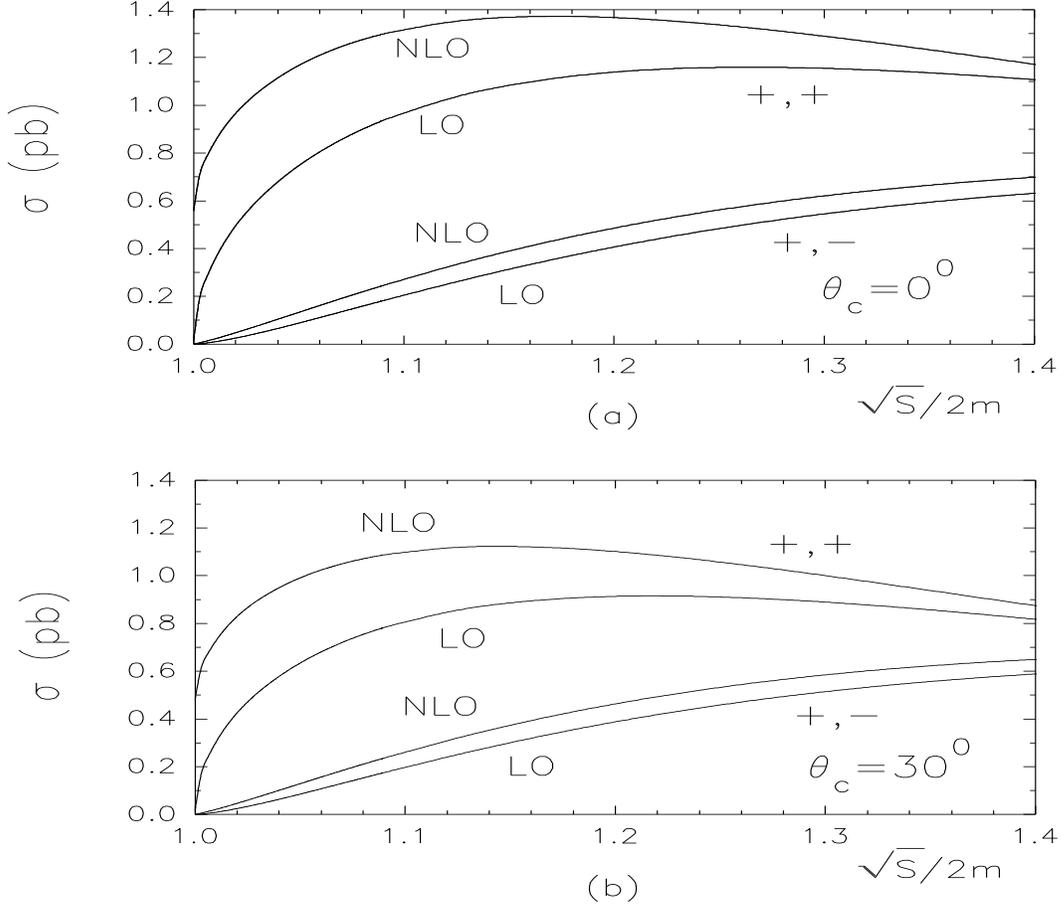}}
\end{center}
\caption{NLO QCD corrected top quark pair production}
\label{fig:f2}
\end{figure}

The strong NLO corrections are large and important in the
region above the threshold up to about 500 GeV. They lead
to a significant increase of the rate in the $(++)$ mode. The angular
cuts do not make a significant difference \cite{kamal} (see Fig.\ref{fig:f2}).

The electroweek corrections are also known \cite{denner}
to be on the order of 0 - (-10)\% up to  1TeV energies.
In the region close to the threshold they are about -10 \%
for the $J_{z}=0$ case. The EW corrections are important for high
accuracy predictions. 

The corrections for the linear polarizations of initial photons
have recently been computed \cite{jikia1}.
\begin{figure}[htbp]
\begin{center}
\mbox{\epsfxsize=14cm\epsfysize=12cm\epsffile{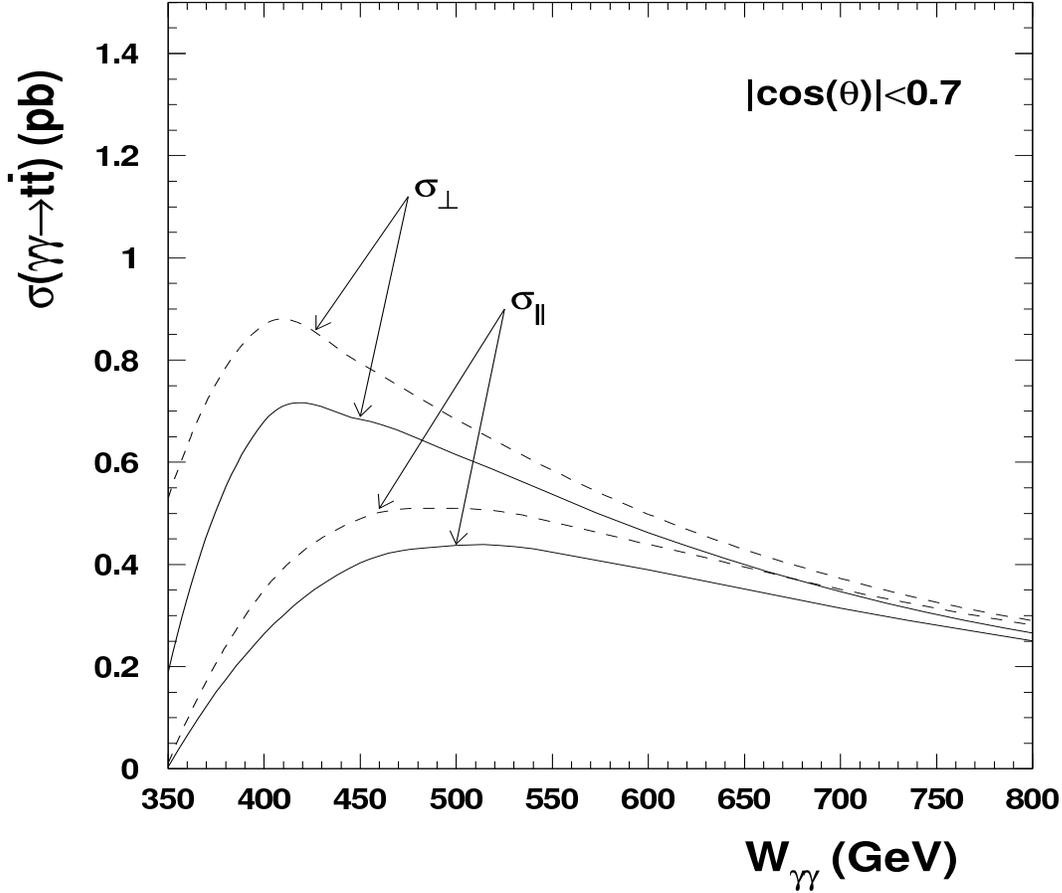}}
\end{center}
\caption{Top quark pair production to NLO for the linear polarizations}
\label{fig:f3}
\end{figure}
The results in Fig. \ref{fig:f3} are given for $\Delta\gamma = 0$ and $\pi/2$ where
$\Delta\gamma$ is the angle between the polarization vectors of colliding
photons. The calculation could be important for a measurement of
CP parity of SUSY Higgs particles H and A.

\section{Top pair production in $\gamma \gamma$ close to the threshold}
Due to the fina state interactions, in order to get a reliable answer
for the production cross section in the threshold region 
it is necessary to resumm the Coulomd corrections.
In this sense the situation is similar to the case of $e^+e^-$ collisions
\cite{ee-thres}.

After a resummation the cross section close to the threshold 
increases 4-5 times. The effect is specially pronounced for
the $(++)$mode  of colliding photons \cite{penin}.
\begin{center}
\begin{figure}[htbp]
\input{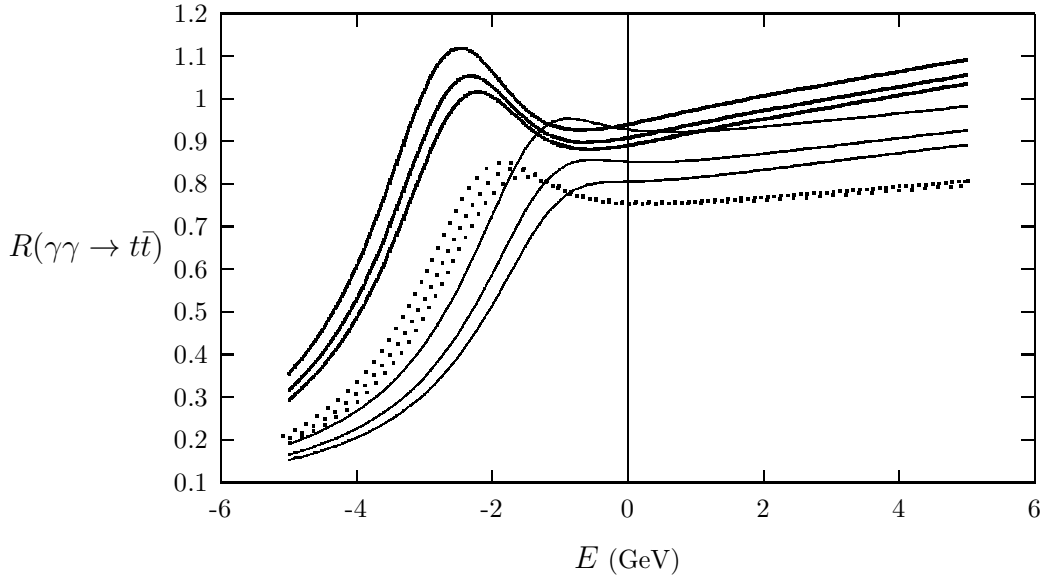}
\caption{The normalized  cross section of top quark pair production 
close to the threshold}
\label{fig:f4}
\end{figure}
\end{center}
In Fig. \ref{fig:f4} the normalized  cross section 
$R(\gamma\gamma\rightarrow t\bar t)=
{\sigma(\gamma\gamma\rightarrow t\bar t)\over
\sigma(e^+e^-\rightarrow\mu^+\mu^-)}$
is shown  
for unpolarized photons as a function of energy
counted from threshold in the leading order (weak solid line),
NLO (dotted line) and
NNLO (bold solid line) for $m_t=175$ GeV  and 
$\alpha_s(M_Z)=0.118$. The ``soft'' normalization scale
for the gluons responsible for the 
Coulomb binding effects is $\mu_s=50~{\rm GeV}$. 
Only the known logarithmic part of the 
``hard'' matching coefficient is used in the NNLO cross section.

However one should note that smearing related to the 
energy spectrum of colliding photons was not included in the 
calculations.

\section{Probe for anomalous couplings from pair production}
There are two points which are different for the case of $\gamma\gamma$ 
and $e^+e^-$ collisions with respect to the couplings:
\begin{itemize}
\item in $\gamma\gamma$ collisions the $\gamma t \bar{t}$ coupling is
involved in the 4th power 
\item the $\gamma t \bar{t}$ coupling is separated from $Z t \bar{t}$
coupling in $\gamma\gamma$ collisions while in $e^+e^-$ collisions both couplings are
involved.
\end{itemize}

One can use the folowing effective Lagrangian which includes
anomalous formfactors in a general model independent way:

$$L_{eff}=ie_{0} ( f_1^{\alpha}\gamma_{\mu} + \frac{i}{2m_{t}}
f_2^{\alpha}\sigma_{\mu\nu}q^{\nu} + f_3^{\alpha}\gamma_{\mu}\gamma_5  
+ \frac{i}{2m_{t}}f_4^{\alpha}\sigma_{\mu\nu}\gamma_5 q^{\nu}), $$
where $\alpha = \gamma,Z$, and, of course, only couplings
with $\alpha = \gamma$ occur in $\gamma\gamma$ collisions. 

\begin{figure}[htbp]
\centering\mbox{\epsfxsize=14cm\epsfysize=12cm\epsffile{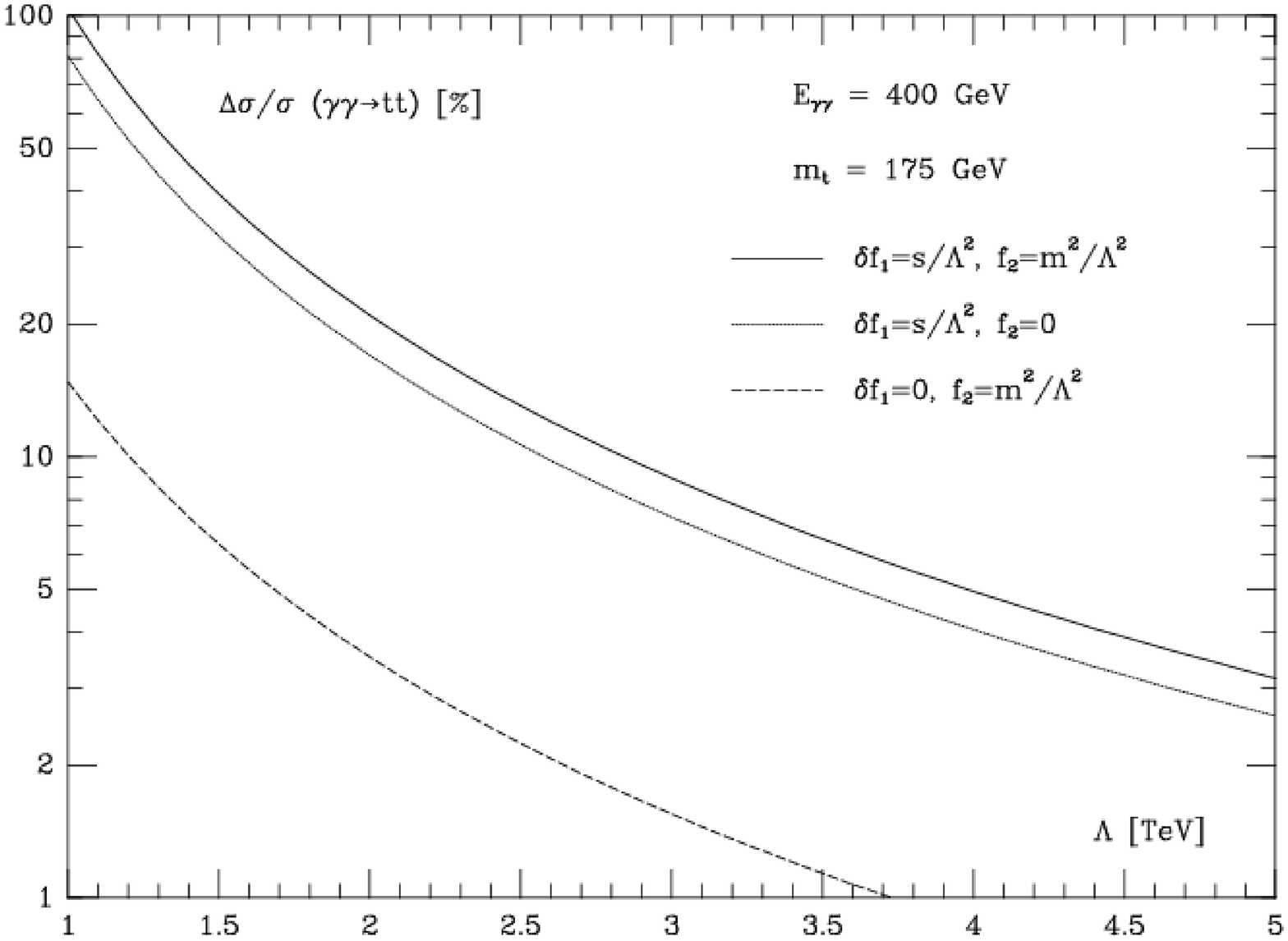}}
\caption{Deviations from the SM cross section in \%}
\label{fig:fig5}
\end{figure}

It was demonstrated \cite{djuadi} that
if one can measure the cross section with 2\% accuracy
one will be able to probe $\Lambda$ upto 10 TeV 
where the formfactors reexpressed through $\Lambda$ as
$f_i^{\alpha} \rightarrow (f_i^{\alpha})^{SM}(1 + s/\Lambda^2)$
Results are shown in Fig.\ref{fig:fig5}.

Sensitivity to the anomalous magnetic moment $f_2^{\gamma}$
is slightly better in $\gamma\gamma$ than in $e^+e^-$
collisions.

The $f_4^{\alpha}$ term is the CP violating term. The best limit on
the imaginary part of the electric dipole moment $Im(f_4^{\gamma})$ is
about $2.3~10^{-17}$ \cite{rindani}. It comes from the forward-backward
asymmetry
$A_{fb}$ with initial-beam helicities of
electron and laser beams $\lambda_e^1=\lambda_e^2$
and $\lambda_l^1= -\lambda_l^2$. The limit for the real part
of the dipole moment is also on the order of  $10^{-17}$
obtained from the linear polarization asymmetries \cite{choi}.
One should stress the limit is an order of magnitude better 
than that obtained from 500 GeV $e^+e^-$ collisions. 

The top quark pair production in $\gamma\gamma$ collisions provides a very
interesting option to study parity properties of the Higgs boson and CP
violating effects \cite{kuehn}. 
It was shown for the two-Higgs doublet model that there are several
asymmetries which are useful for probing CP violation in the 
Higgs sector \cite{anlauf}.

The problem here is that there is an interference between the H and the A with
a small mass gap.
A model-independent study of the effects of a neutral Higgs bosons 
without definite CP-parity has been done \cite{asakawa} where
complete set of asymmetries was proposed in order to determine
all the CP couplings. 
However, the effect is
less pronounced with increasing $tan\beta$. An expected mass
resolution for the top quark is about 5-10 GeV,
and therefore it is problematic to separate signals 
for various polarization configurations from the background \cite{zerwas}.

The $\gamma\gamma$ colliders will be comparable to the
$e^+e^-$ and $e^-e^-$ in the search reach for large extra
dimensions via top pair production \cite{rizzo}.

\section{Single top production in $\gamma \gamma$ and $\gamma e$}
\begin{figure}[htbp]
\centering\mbox{\epsfxsize=8cm\epsfysize=8cm\epsffile{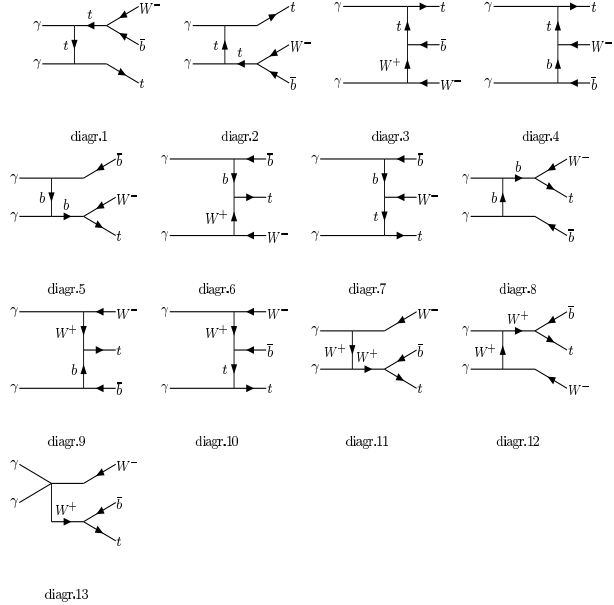}}
\caption{Complete set of SM tree diagrams for $W t b$ production}
\label{fig:f6}
\end{figure}

\begin{figure}[htbp]
\centering\mbox{\epsfxsize=10cm\epsfysize=10cm\epsffile{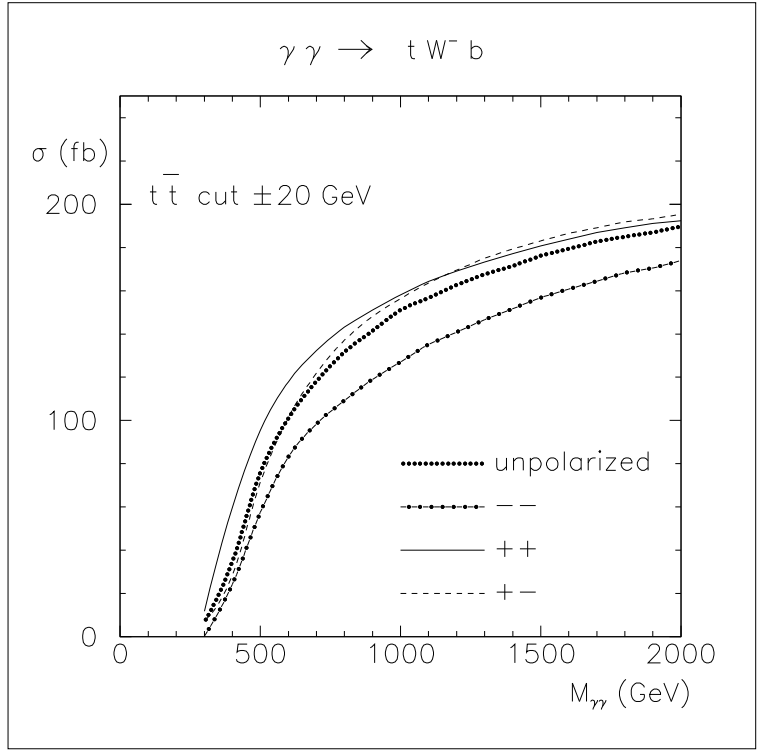}}
\caption{Single top production rate in $\gamma\gamma$ collisions for various
polarizations}
\label{fig:f7}
\end{figure}

Single top production in $\gamma\gamma$ collisions results in the same final state
as the top pair production. Here the situation 
is similar to single top production at the the LHC in $Wt$ mode \cite{belyaev-boos}.
The top pair production rate must 
be removed from the total $\gamma \gamma \rightarrow W t b$ rate
(see the complete set of tree diagrams in Fig.\ref{fig:f6})in
order to get the correct single top production rate. This should be done in a gauge
invariant way. One can make the fit of the peak in the $Wb$
invariant mass with the Breit-Wiegner formula and then remove the
peak contribution, or one can apply a cut of $\pm 25$ GeV around
the peak position in the $Wb$ invariant mass distribution.
These two methods lead to very similar numerical results
\cite{boos-gg} presented in the Fig.\ref{fig:f7}. 
\footnote{All the computations have been done by means of the
program CompHEP \cite{comphep}}

\begin{figure}[htbp]
\centering\mbox{\epsfxsize=10cm\epsfysize=4cm\epsffile{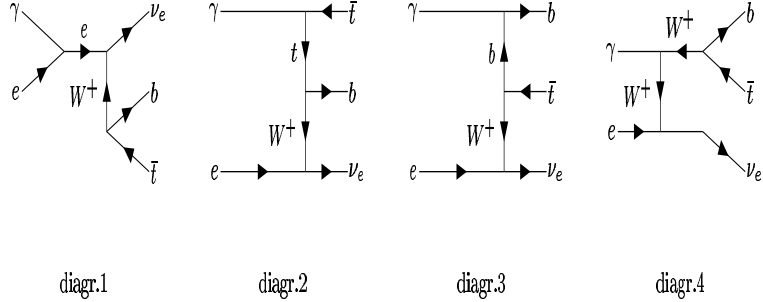}}
\caption{SM tree diagrams for single top production in $\gamma e$
collisions}
\label{fig:f8}
\end{figure}

Single top production in $\gamma e$ collisions has been discussed
in several papers \cite{singletop}. The complete set of SM tree
diagrams is shown in Fig. \ref{fig:f8}. 
In contrast to the top pair production rate the single top rate
is directly proportional to the $Wtb$ coupling and therefore 
the process is very sensitive to its structure.

In terms of the notation as from the effective
Lagrangian
\begin{eqnarray}
{L} & \left.=  \frac{g}{\sqrt{2}}\right[ &
 W_{\mu}^-\bar{b}(\gamma_{\mu}f_{1L} P_- +
\gamma_{\mu} f_{1R} P_+) t \nonumber \\
 & & - \left.\frac{1}{2M_W} W_{\mu\nu}
\bar{b}\sigma^{\mu\nu}(f_{2R} P_- + f_{2L} P_+) t
\right] + {\rm h.c.},
\label{eq:lagrangian_anom} 
\end{eqnarray}
where $W_{\mu\nu} =
D_{\mu}W_{\nu} - D_{\nu}W_{\mu}, D_{\mu} = \partial_{\mu} - i e A_{\mu},
P_{\pm} = 1/2(1 \pm \gamma_5)$ and \mbox{$\sigma^{\mu\nu}
= i/2(\gamma_{\mu}\gamma_{\nu} - \gamma_{\nu}\gamma_{\mu})$}
one can compare collider potentials for an expected accuracy
of the anomalous parameter measurements presented in the table \ref{tb:par}.
The notations are related to those from other studies \cite{gounaris}
by the formula
$f_{2L(R)} = \frac{C_{t(b)W \Phi}}{\Lambda^2}\frac{v\sqrt{2} \, m_W}{g}$,
where $\Lambda$ is the scale of
new physics.

\begin{table}
  \begin{center}
    \begin{tabular}{|l|c|c|}\hline
        &{$F_2^L$}
        &{$F_2^R$} \\\hline\hline
      Tevatron ($\Delta_{sys.}\approx10\%$)
               & $-0.18$ $\div$$+0.55$ & $-0.24$ $\div$$+0.25$ \\ 
      LHC ($\Delta_{sys.}\approx5\%$)
               & $-0.052$$\div$$+0.097$ & $-0.12$ $\div$$+0.13$ \\
      $\gamma e$ ($\sqrt{s_{e^+e^-}}=0.5$ TeV)
               & $-0.1$ $\div$$+0.1$ & $-0.1$ $\div$$+0.1$ \\  
      $\gamma e$ ($\sqrt{s_{e^+e^-}}=2.0$ TeV)
               & $-0.008$$\div$$+0.035$ & $-0.016$$\div$$+0.016$
\\
      \hline
    \end{tabular}%
  \end{center}
\caption{Expected sensitivity for Wtb anomalous couplings measurements} 
\label{tb:par}
\end{table}

\begin{figure}[htbp]
\centering\mbox{\epsfxsize=10cm\epsfysize=10cm\epsffile{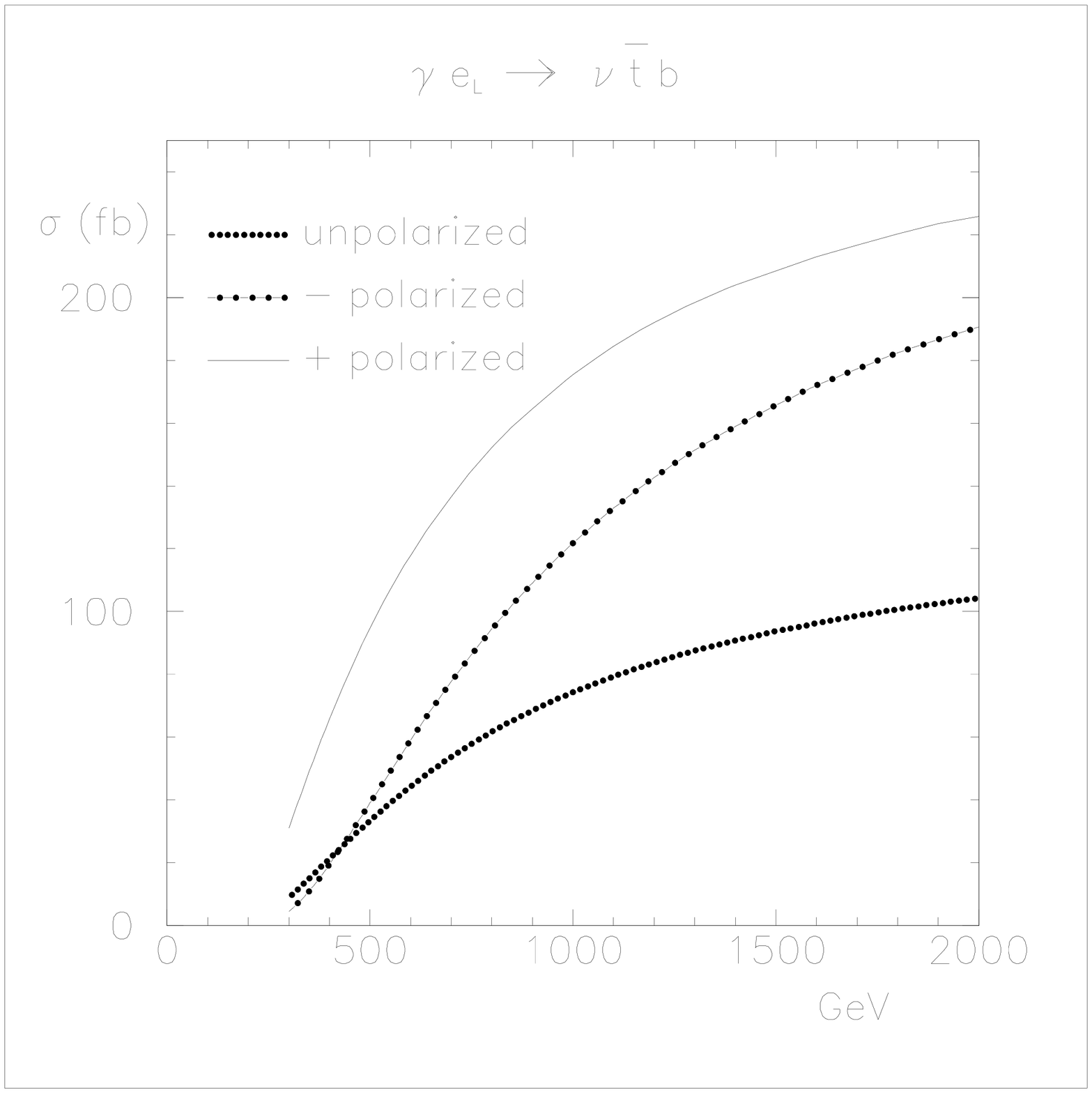}}
\caption{Single top quark production cross section in $\gamma e$ 
collisions}
\label{fig:f9}
\end{figure}

In the table \ref{tb:par} \cite{boos-dudko-ohl} uncorrelated limits on anomalous
couplings from
measurements at different machines are shown. One can see
the best limits one can reach at very high energy 
$\gamma e$ colliders even in the case of unpolarize collisions.
In the case of polarized collisions, the rate is increasing
significantly as shown in Fig.\ref{fig:f9} \cite{boos-gg} and therefore one can get
better bounds. One should stress that only left handed electrons lead to
a nonvanishing cross section, whereas the cross section
with righ-handedt electrons is proportional to the electron
mass squared and therefore nigligible.

Single top quark production in 
$\gamma \gamma$ and $\gamma e$ collisions provides interesting
possibilities to test FCNC radiative couplings and to
study various predictions of Technicolor models.
The following effective Lagrangian is used to parametrize 
anomalous couplings:
$$ L_{eff} = \frac{e}{\Lambda} (k_c \bar{t}\sigma{\mu\nu}c) F^{\mu\nu}
+h.c.~.$$
$k/\Lambda$ is expected to be constrained at the level of 
0.12/TeV at the Tevatron with 10 $fb^{-1}$ and 0.01/TeV at the LHC
with 100 $fb^{-1}$ integrated luminosity.
At a 500 GeV $\gamma\gamma$ collider one expects a limit for $k/\Lambda$ 
about  0.05/TeV with 10 $fb^{-1}$ luminosity \cite{abraham}.

\begin{figure}[htbp]
\centering\mbox{\epsfxsize=10cm\epsfysize=10cm\epsffile{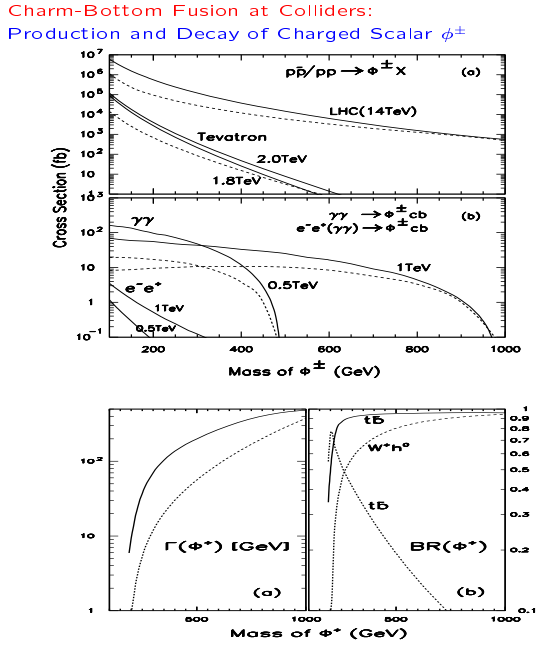}}
\caption{Production and decay of the Technicolor charged scalar} 
\label{fig:f10}
\end{figure}

There are many variants of technicolor models. The detailed predictions
are normally model dependent. However, there are predictions, like
that of the existence of charged (pseudo-)scalars, which are somewhat model
independent. Several studies \cite{wang, he} of that have been done for photon
colliders. If there is a large flavor mixing between the right-handed top
and charm quarks it leads to a large Yukawa coupling of a charged 
(pseudo-)scalar with charm and bottom quarks. The dominant decay mode
in Topcolor models is $\phi^+ \to t\bar{b}$. So it gives the contribution
to the single top production in $\gamma\gamma$ and $\gamma e$ collisions
\cite{he} (see results in Fig.\ref{fig:f10}).

\vspace{-1cm}
\section{Concluding Remarks and Problems}
In this short review we have considered various aspects of 
top quark physics which can be studied at linear $\gamma\gamma$
and $\gamma e$ colliders. 
\begin{itemize}
\item{Top quarks could be produced at photon colliders in the top pair
mode or singly.  The SM production mechanisms are clean and simple.
The SM cross section and distributions are known
to the LO level and in several cases to the NLO level.}
\item{Photon Linear Colliders provide a number of 
interesting options which are comparable and in some cases even better than
those at other colliders
for the study of top quark physics.}
\begin{itemize}
\item{The top pair production rate is proportional to the coupling
of a top quark with a photon in 4th power and is therefore very sensitive 
to its structure and possible deviations from the SM.}
\item{Using various asymmetries one can uniquely study 
CP properties of the Higgs bosons and CP violating effective
operators}.
\item{Single top quark production at a high energy $\gamma e$
collider is the best collision option for the study of the
structure of the $Wtb$ coupling}.
\item{Single top production at photon colliders allows one to study various
effects predicted by Technicolor models} 
\end{itemize}
\item{One should stress that in general, simulations
of top quarks effects at photon colliders are much less
developed than for $e^+e^-$ LC and the hadron colliders 
Tevatron and LHC. Usually subsequent decays, jet fragmentation,
a possible detector response, as well as various backgrounds 
have been not included in the simulations. An important 
problem is to simulate correctly the influence of a real
photon spectrum. Much work still
needs to be done.} 
\end{itemize}
\vspace{-0.3cm}
We must appologize that not all
the results obtained on the subject have been mentioned.
The author thanks the Organizing Committee of the
Workshop for kind hospitality and financial support.
The work was partly supported by the RFBR-DFG 99-02-04011,
RFBR 00-01-00704, and CERN-INTAS 99-377 grants.

\end{document}